\documentclass[a4paper,12pt]{article}

\usepackage{amsmath,amssymb,latexsym}
\usepackage[T1]{fontenc}
\usepackage{graphicx}
\usepackage{makeidx}
\usepackage{slashed}
\usepackage{color}
\usepackage[centering]{geometry}

\begin{document}
\vspace*{-3cm}
\begin{flushright}
  Freiburg-PHENO-09/02
\end{flushright}
\vspace*{1.5cm}

\begin{center}
  {\LARGE{Angular correlations in $t$-channel single top production at the LHC}}\\
  \vspace*{1cm}
  Patrick Motylinski\\\footnotesize{\verb+patrick.motylinski@physik.uni-freiburg.de+}\\
  {\footnotesize{Physikalisches Institut, Albert-Ludwigs-Universit\"at Freiburg\\Hermann-Herder-Stra\ss e 3, D-79104 Freiburg i.Br., Germany}}
\end{center}
\vspace*{0.25cm}


\begin{abstract}
\noindent When a top quark decays there is a large amount of angular correlation, in its rest frame, between its spin orientation and the direction of flight of the charged lepton from its decay. In this letter we investigate the prospects of measuring this angular correlation using the MC@NLO framework. The strength of the correlation is investigated for different spin bases. The robustness against variations of PDF sets and uncertainties, factorization scale dependence, center-of-mass energy, and the jet $R$--parameter, is also examined. 
\end{abstract}

\section{Introduction}
\label{sec:intro}

Due to its very large mass the relevant time scale for the decay of a top quark is so short that QCD interactions do not alter the spin orientation. Consequently the top quark decays with the spin it was born with. This makes single top production a valuable source of information when studying the coupling between a fermion and a vector boson. In order to extract this information, however, suitable observables must be identified. The angular correlation between the spin orientation of the top and the charged lepton from its decay is such an observable. The strength of such correlations depends on the degree of polarization of the top quarks. By means of single top production it is possible to collect samples of highly polarized top quarks. The amount of polarization depends, however, on the choice of the top spin quantization axis. As we shall see, it is possible to choose spin bases in which the top quarks are strongly polarized.\\ 
The $t$-channel (sometimes called the $Wg$-fusion channel) is predicted to be dominant at the LHC, with a cross section~$\sim$250 pb, compared to the $s$-channel process with~$\sim$15 pb
and the $Wt$ associated production with~$\sim$60 pb. In $Wt$ associated production the single top is produced together with the $W$-boson with the dominant initial-state configuration being $bg$. The study of angular correlations in $Wt$ associated production is challenging due to the presence of interferences (at next-to-leading order (NLO) in the $Wt$ case) with certain $t\bar{t}$ final-states~\cite{Boos:2002xw,Frixione:2008yi}.\\
Measuring the $s$-channel production mode at the LHC will be very difficult due to its small cross section. As we shall discuss the angular correlations require identifying the direction of isospin $-1/2$ fermions ($d$-type quarks and charged leptons). The measurement of angular correlations in the $s$-channel is cumbered by the fact that the LHC is a proton--proton collider which means that neither of the incoming beams provide a copious source of $\bar{d}$-quarks. This makes it more difficult to choose a basis in which the top quarks are strongly polarized~\cite{Bernreuther:2008ju}.\\
Experimentally, it will be challenging to measure the angular correlations. The kinematical frames in which the correlations are measured (incoming parton center-of-mass frame, top rest frame etc.) will have to be reconstructed on an event-by-event basis, using the available information (energies and 3-momenta of the charged lepton and spectator jet, energy conservation in the transverse plane) as well as assumptions regarding the decay of the top (on-shell decay or using a decay width) and the top mass. Uncertainties will therefore be playing a significant role and predictions of high precision are a necessity.\\
In this letter we therefore focus on $t$-channel production. We constrain ourselves to investigating the angular correlations in the top rest frame and we do not take any detector effects into account.
In our study of the angular correlations we have used the MC@NLO framework~\cite{Frixione:2002ik}. Single top $s$- and $t$-channel production with the inclusion of angular correlations were included in MC@NLO some time ago~\cite{Frixione:2005vw,Frixione:2007zp}, making MC@NLO suitable for our study. The results presented here are thus accurate to NLO\footnote{Virtual corrections are not included in the decay of the top.}, both with respect to real and virtual corrections, using leptonic decays of the top/anti-top\footnote{By leptonic decay we here mean the following: $t \rightarrow b + W\{\rightarrow l^+ + \nu \}$}, and hadronization effects are also taken into account. While a study of angular correlations in $t$-channel single top production including detector effects has been performed (see~\cite{Beneke:2000hk}) ours uses an NLO event generator to accurately assess how the angular correlations depend on various parameters relevant for the LHC.\\
We start out by reviewing two different choices of spin quantization bases in Section~\ref{sec:theory}
which will be based on the definitions given in~\cite{Mahlon:1999gz}. 
We then proceed to present numerical results in Section~\ref{sec:results}. In particular we shall look at single top and anti-top production separately as these production modes are not symmetric at the LHC. We also investigate the robustness of the correlations against changing the parton distributions functions (PDF's), varying the factorization scale and taking PDF errors into consideration, lowering the center-of-mass energy and varying the $R$--parameter of the spectator jet. The results are summarized in tables given at the end of this letter. Finally, we give our conclusions in Section~\ref{sec:concl}.

\section{Angular Correlations in the $t$-channel}
\label{sec:theory}

We consider a situation where the top quarks can be in a spin up or a spin down state and assume that the top decays into a final state consisting of a $b$-quark, a charged lepton and a neutrino (the latter two are decay products of the intermediate $W$-boson). The differential decay rate then becomes~\cite{Mahlon:2000ze}:
\begin{equation}
  \label{eq:decay1}
  \frac{1}{\Gamma_t}\frac{d \Gamma}{d \cos \theta} = \frac{1}{2}(1+ \mathcal{A}_{\uparrow \downarrow} \cdot s \cdot \cos \theta)
\end{equation}
where $\theta$ is the angle between the charged lepton and the top spin quantization axis in the top rest frame, $\Gamma_t$ is the total decay width of the top, $s$ is the correlation coefficient, $\mathcal{A}_{\uparrow \downarrow} \equiv (N_{\uparrow}-N_{\downarrow})/(N_{\uparrow}+N_{\downarrow})$ is the spin asymmetry and $N_{\uparrow / \downarrow}$ is the number of spin up-/down top quarks. The correlation coefficient $s$ is 1 for charged leptons and $\bar{d}$-type quarks ($d$-type for anti-top production), i.e. there is 100\% correlation in those cases. The spin asymmetry is basis dependent with $\mathcal{A}_{\uparrow \downarrow}=1$ in a basis where the top quarks would be 100\% polarized. The fact that the $\bar{d}$-/$d$-type quarks are maximally correlated with the top/anti-top quarks in top/anti-top decays implies, by crossing symmetry, that the $\bar{d}$-/$d$-type quarks present in either initial- or final-state in single top production also are maximally correlated to the top. This is the reason why the spectator basis and the beamline basis, which are introduced and described in large detail in~\cite{Mahlon:1999gz}, are good candidates for top spin quantization bases. In the following we give a brief introduction to the two bases. \\

\subsection{Spectator Basis}
\label{sec:spectator}
\begin{figure}[h!]
  \centering
  \includegraphics[height=3cm]{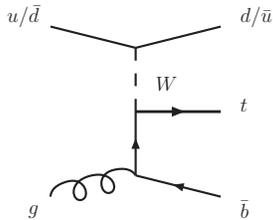}
  \caption{Single top $t$-channel diagram.}
  \label{fig:t-chan}
\end{figure}
\noindent Prior to the decay of the top the final-state in $t$-channel single top production consists of top, a $b$-quark (from gluon-splitting in the initial-state) and a spectator quark/jet (see Fig.~\ref{fig:t-chan}). In about 80\% of the time the spectator quark is a $d$-type, mainly due to the large abundance of valence $u$-type quarks in the initial state. Thus, choosing the direction of the spectator jet as quantization axis ensures that the top quarks will be strongly polarized.\\
In the case of anti-top production the situation is somewhat different. There
is only a $d$-type quark in the final-state about 31\% of the time, which suggests that choosing the
spectator direction is not an appropriate choice in the case of anti-top quark production. However
this is not quite the case. The spectator quark is essentially a quark being
scattered off a $W$-boson and the deflection is rather limited, i.e. the incoming $d$-type quark is
nearly aligned with the outgoing spectator. Therefore the spectator basis
still provides a basis in which there should be a large degree of polarization and, therefore, angular correlations. We shall see this confirmed in the next section by numerical results.

\subsection{Beamline Basis}
\label{sec:beaml}
\
The dominant initial state for single anti-top production is one with a
$d$-type quark. Therefore, choosing the direction
of the beamline containing the $d$-type quark as top spin quantization axis
provides a highly polarized sample of top quarks and strong angular correlations between the direction of the beamline and the direction of flight of the charged lepton. 
Since the spectator quark/jet in general is only deflected a little from the direction of the incoming beam we can use the pseudo-rapidity of the spectator-jet to determine which beamline to choose as quantization axis. In order to determine the beam providing the $d$-type quark that shares the vertex with the spectator jet, a selection criteria is applied, using the spectator jet pseudorapidity\footnote{The beams are, in general, not back-to-back in the top quark rest frame.}: if $\eta_{j_1} > 0$ then choose the right-moving beam as spin axis and if $\eta_{j_1} < 0$ then choose the left-moving beam. \\
The beamline basis provides a highly polarized sample of anti-top quarks~\cite{Mahlon:1999gz}. It is, however, also suitable for top quarks where the degree of polarization is almost as high. This will be confirmed numerically in the next section. 

\section{Numerical Results}
\label{sec:results}
With the definitions in the previous section we are now able to present
results produced by using MC@NLO. We divide our results in two main parts. The first comprises the results (which we here call the default results) for single top and anti-top production at the
LHC, i.e. $pp$--collisions at $\sqrt{S}=14$ TeV. The predictions are obtained using
the CTEQ66 PDF set~\cite{Nadolsky:2008zw} and as parameters we used $m_t=173.1$~GeV,
$\Gamma_t=1.4$~GeV. The spectator jet is established using the $k_T$--clustering algorithm~\cite{Catani:1993hr}, with
$d_{ \textrm{cut} } = 100$~GeV$^2$. The clustering includes all stable final-state
hadrons and photons. It is assumed that the top decays into a $b$-quark and $W$-boson, the latter decaying into a charged lepton (an $e$ or a $\mu$), a neutrino, and a $b$-quark. For each choice of basis we compare single top and anti-top production.\\
In the second part we investigate the effects of 1) changing the PDF's, varying the factorization scale ($\mu_F$) and assessing the impact of PDF errors, 2) varying the center-of-mass energy of the colliding hadrons and 3) varying the $R$--parameter of the jet (using the $k_T$-clustering algorithm), which amounts to varying the thickness of the jet. The results in the second part are shown for single top production only, being dominant over anti-top production at the LHC.\\

\begin{figure}[t!]
  \centering
  \includegraphics[width=0.49\textwidth]{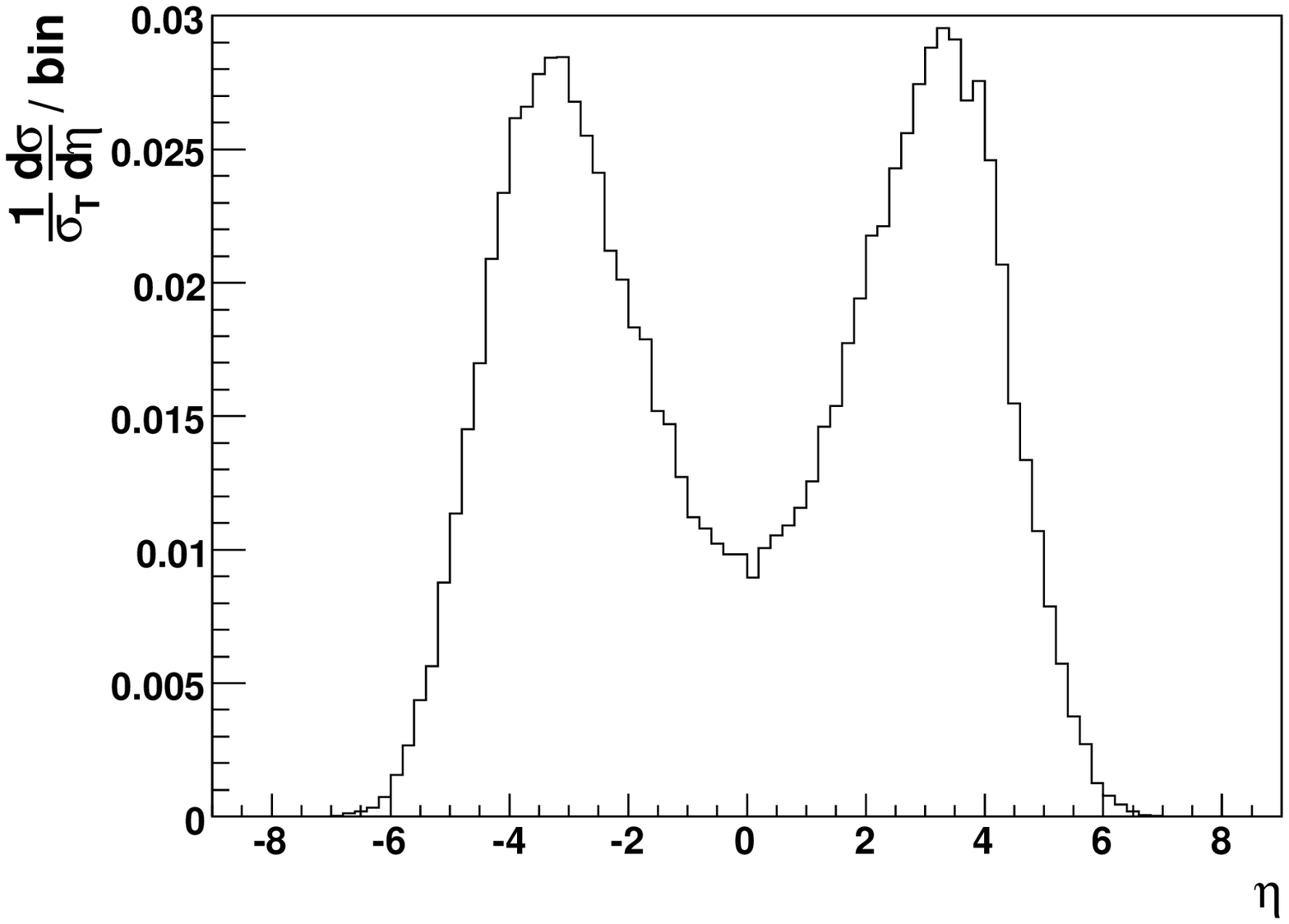}
  \includegraphics[width=0.49\textwidth]{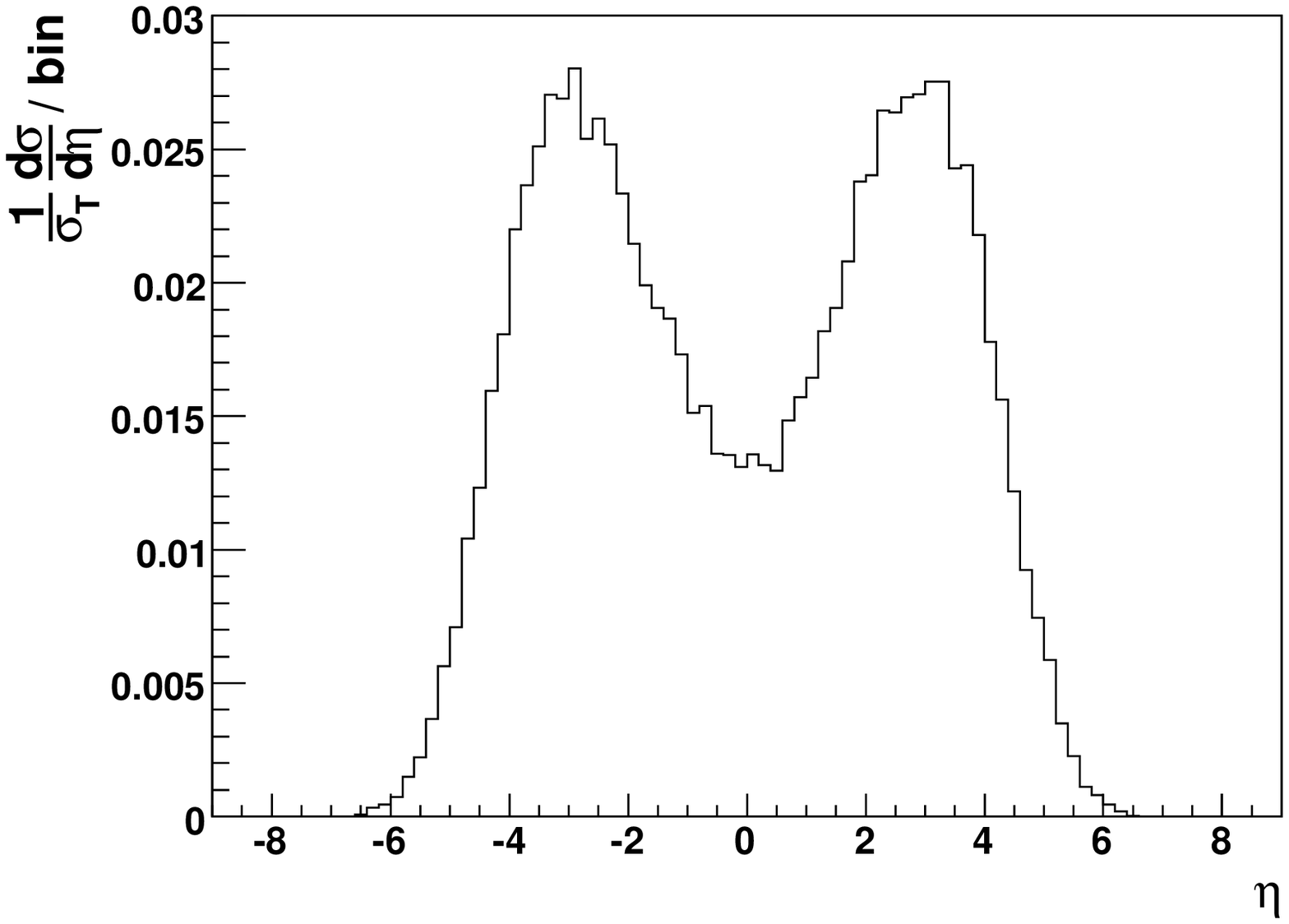}
  \caption{Pseudorapidity of the hardest (non-$b$) jet for single top (left) and single anti-top production (right).}
  \label{fig:jet1-rap}
\end{figure}

\begin{figure}[h]
  \centering
  \includegraphics[width=0.49\textwidth]{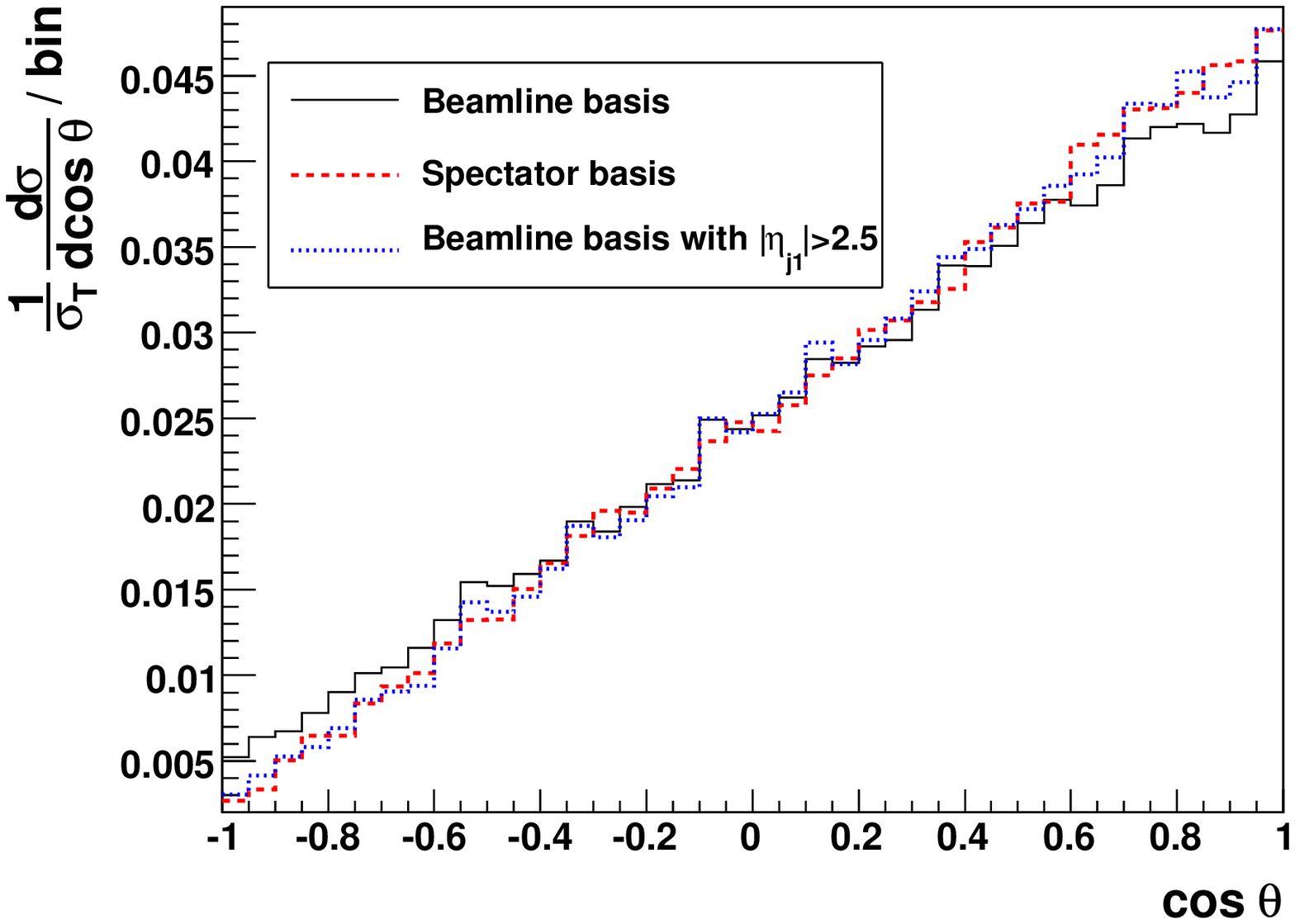}
  \includegraphics[width=0.49\textwidth]{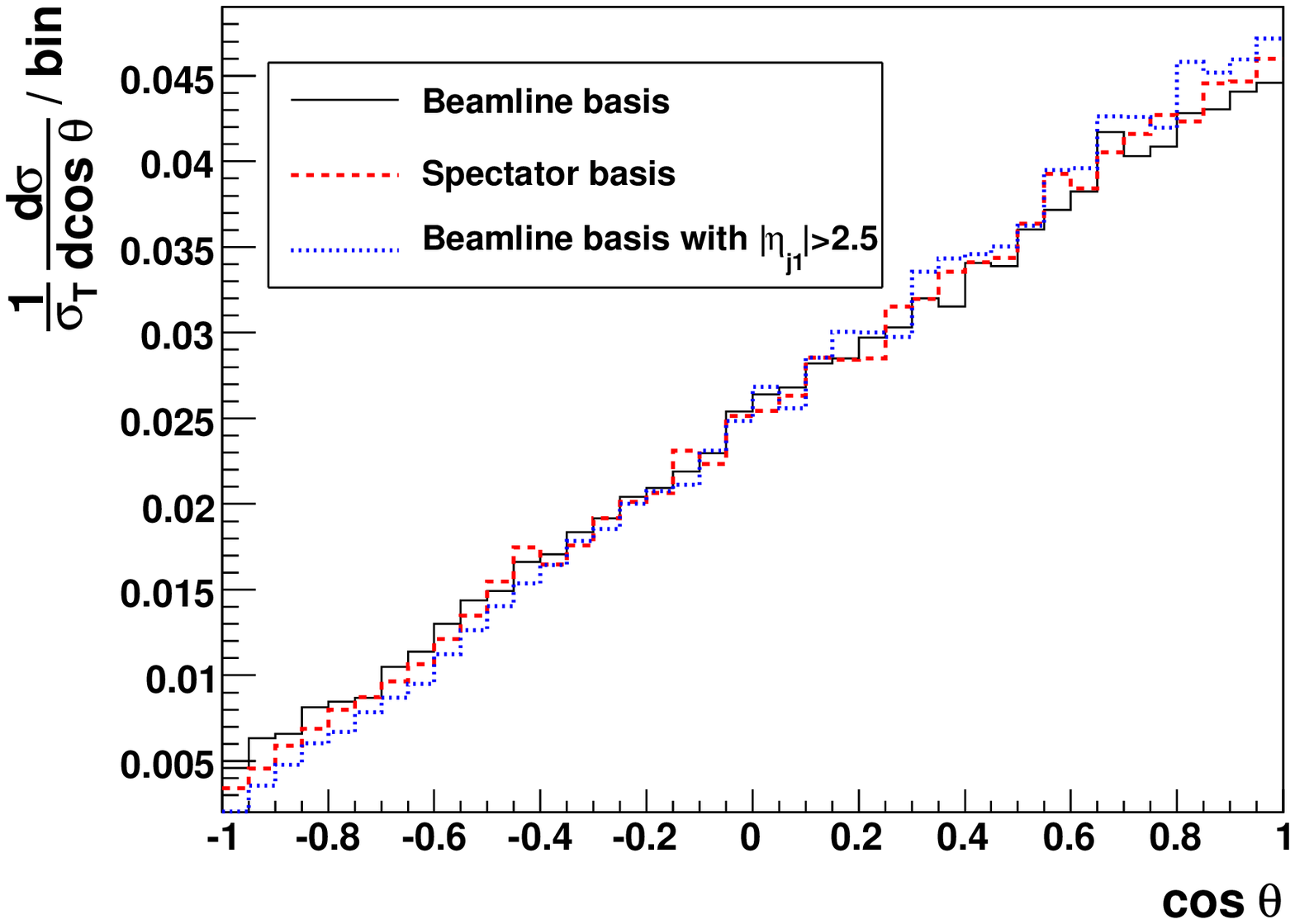}
  \caption{Angular correlation in single top (left) and single anti-top (right) production.}
  \label{fig:correl1}
\end{figure}

\begin{figure}[h!]
  \centering
  \includegraphics[width=0.49\textwidth]{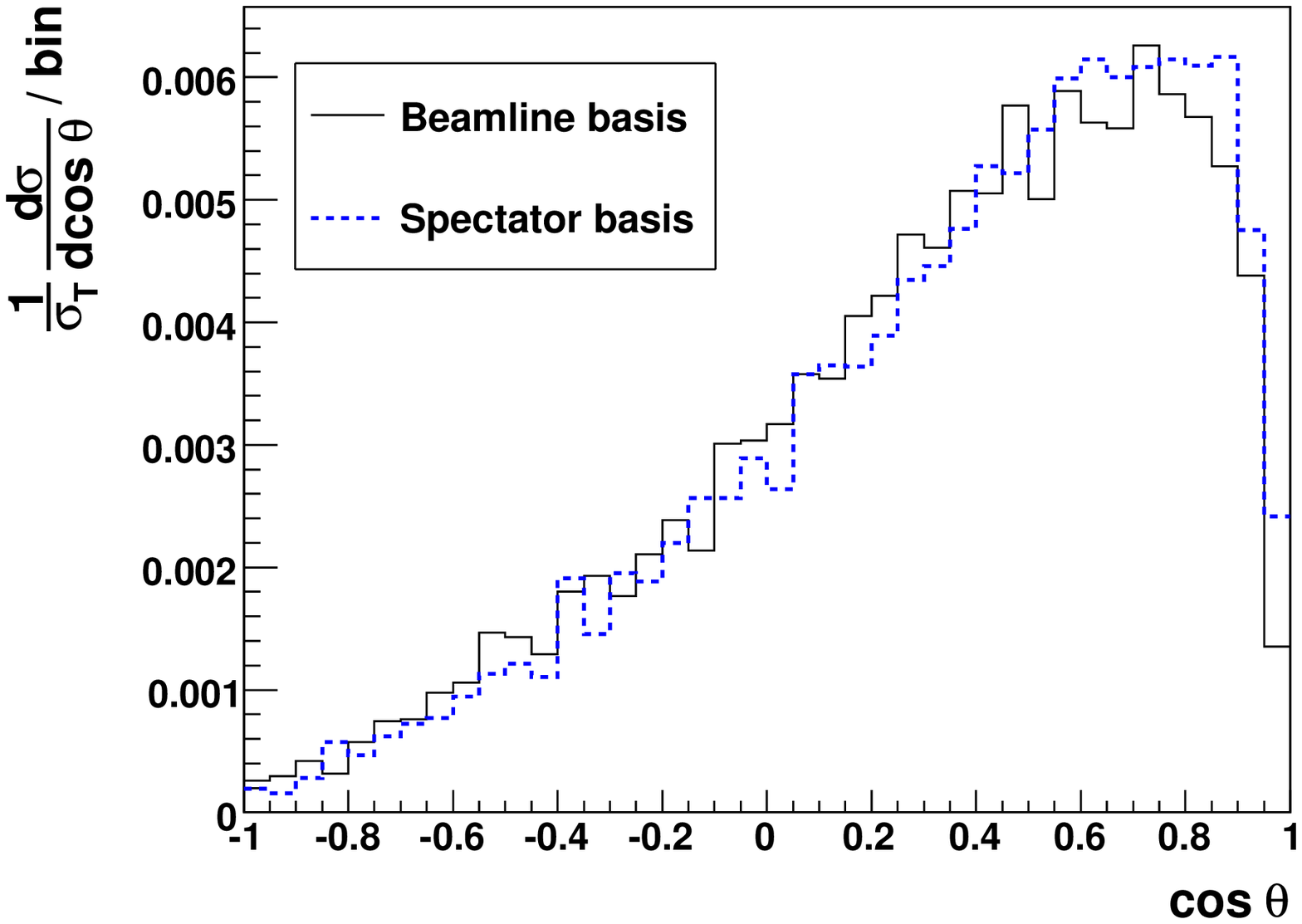}
  \includegraphics[width=0.49\textwidth]{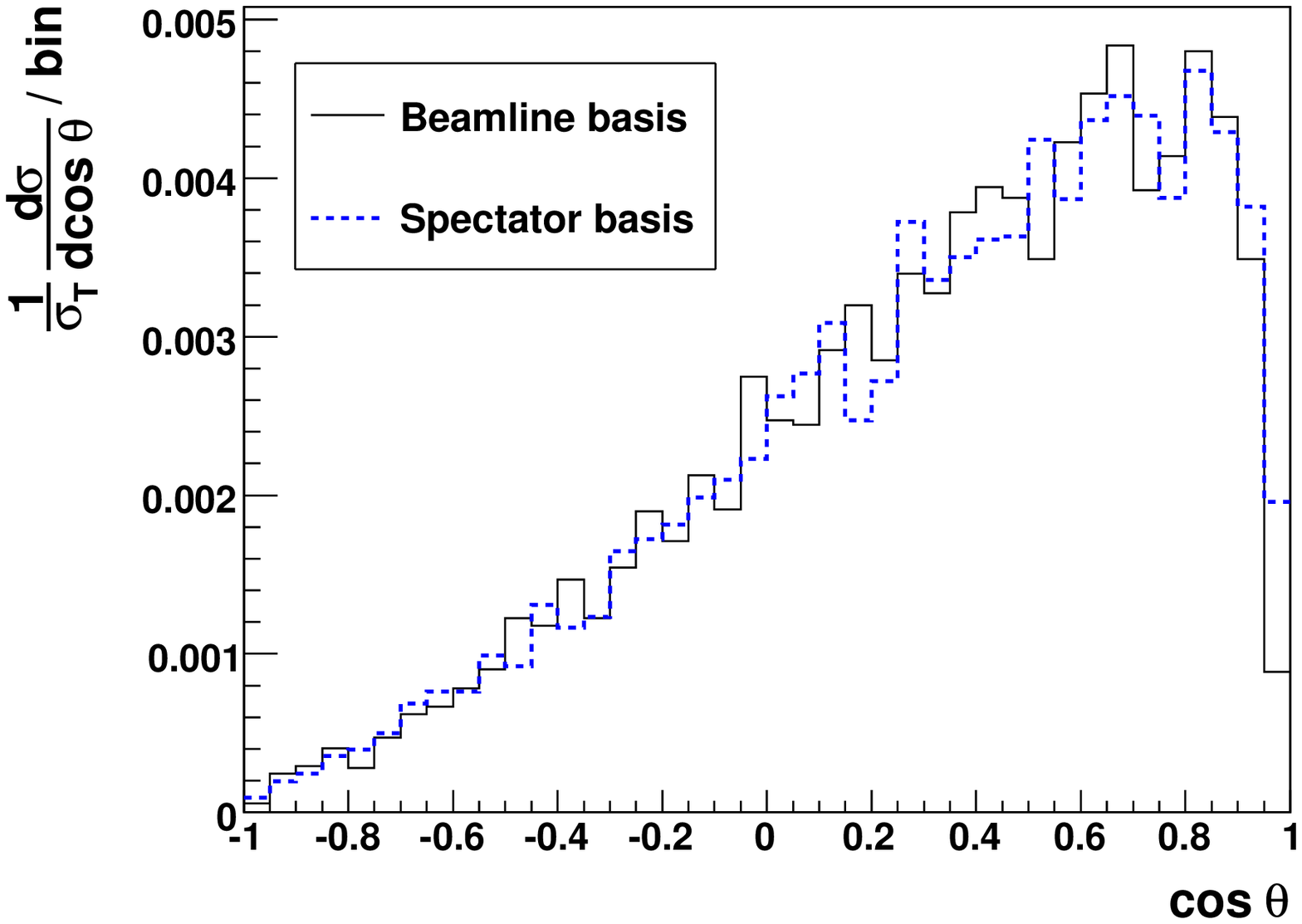}
  \caption{Angular correlation in single top (left) and single anti-top (right) production when cuts are applied (see text).}
  \label{fig:correl2}
\end{figure}

\begin{figure}[t!]
  \centering
  \includegraphics[width=0.49\textwidth]{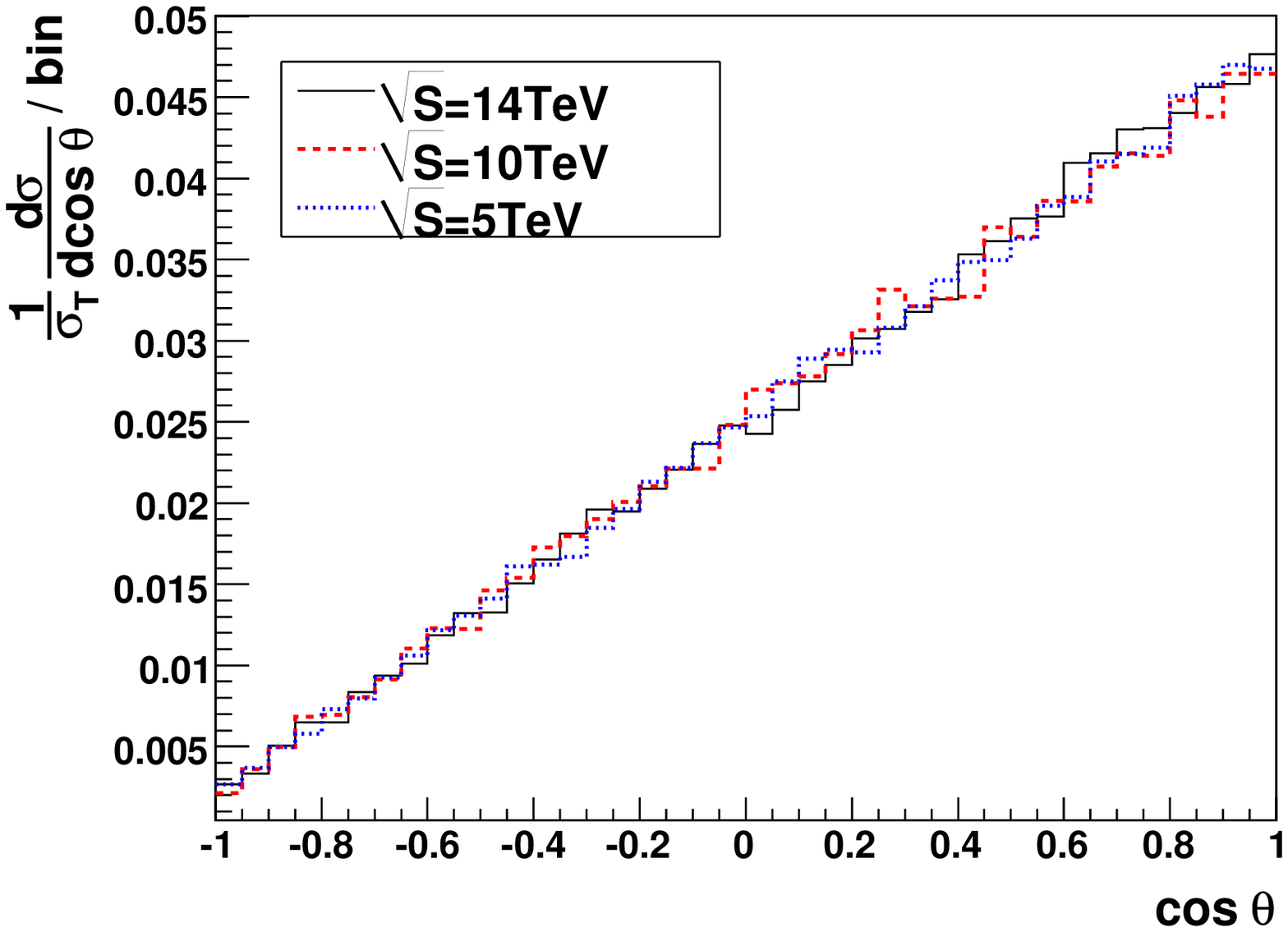}
  \includegraphics[width=0.49\textwidth]{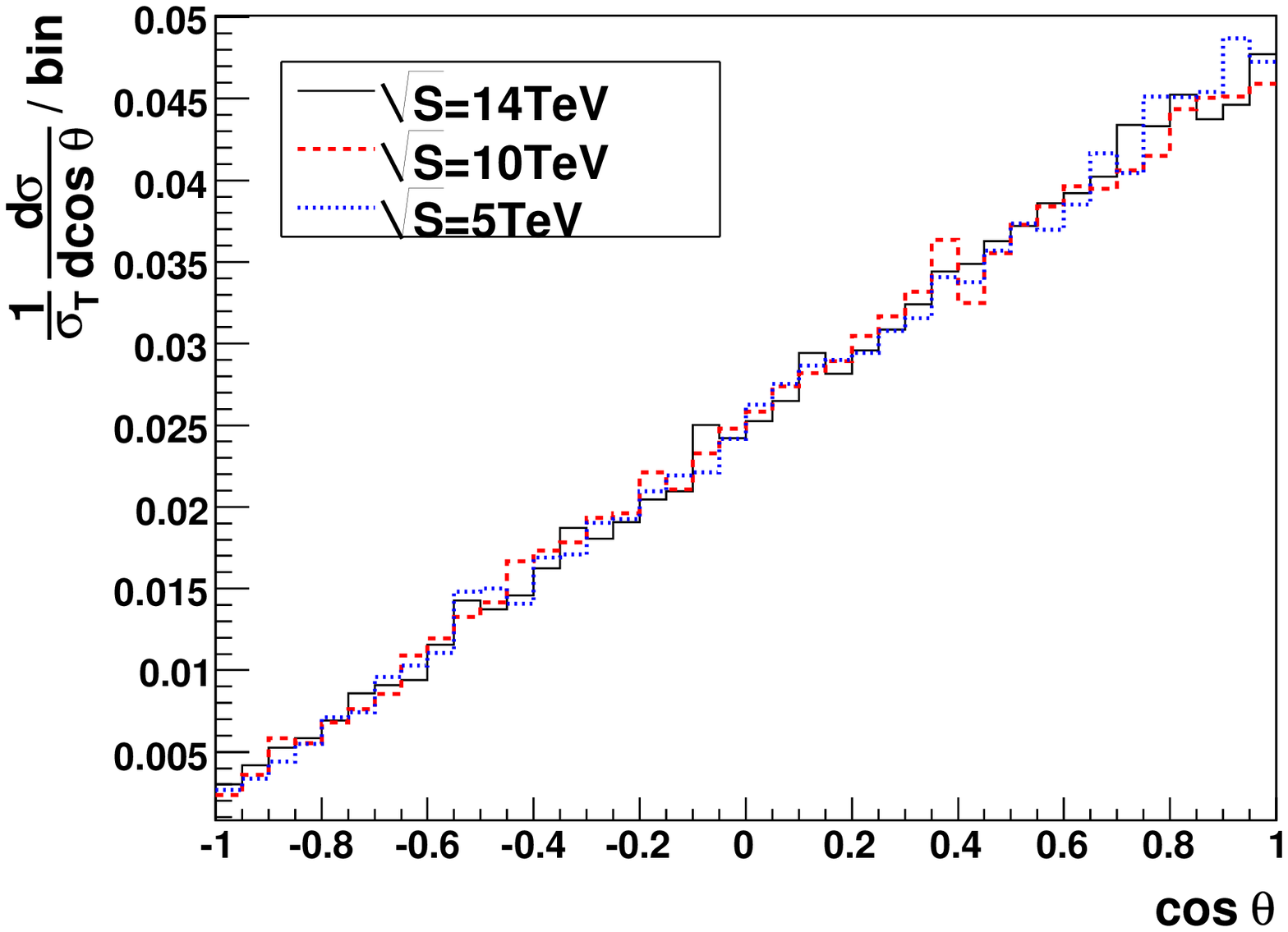}
  \caption{Angular correlation for three different values of the center-of-mass energy. The left plot is for the spectator basis while the right plot is for the beamline basis with $|\eta_{j_1}|>$2.5.}
  \label{fig:ECM}
\end{figure}

\begin{figure}[t]
  \centering
  \includegraphics[width=0.49\textwidth]{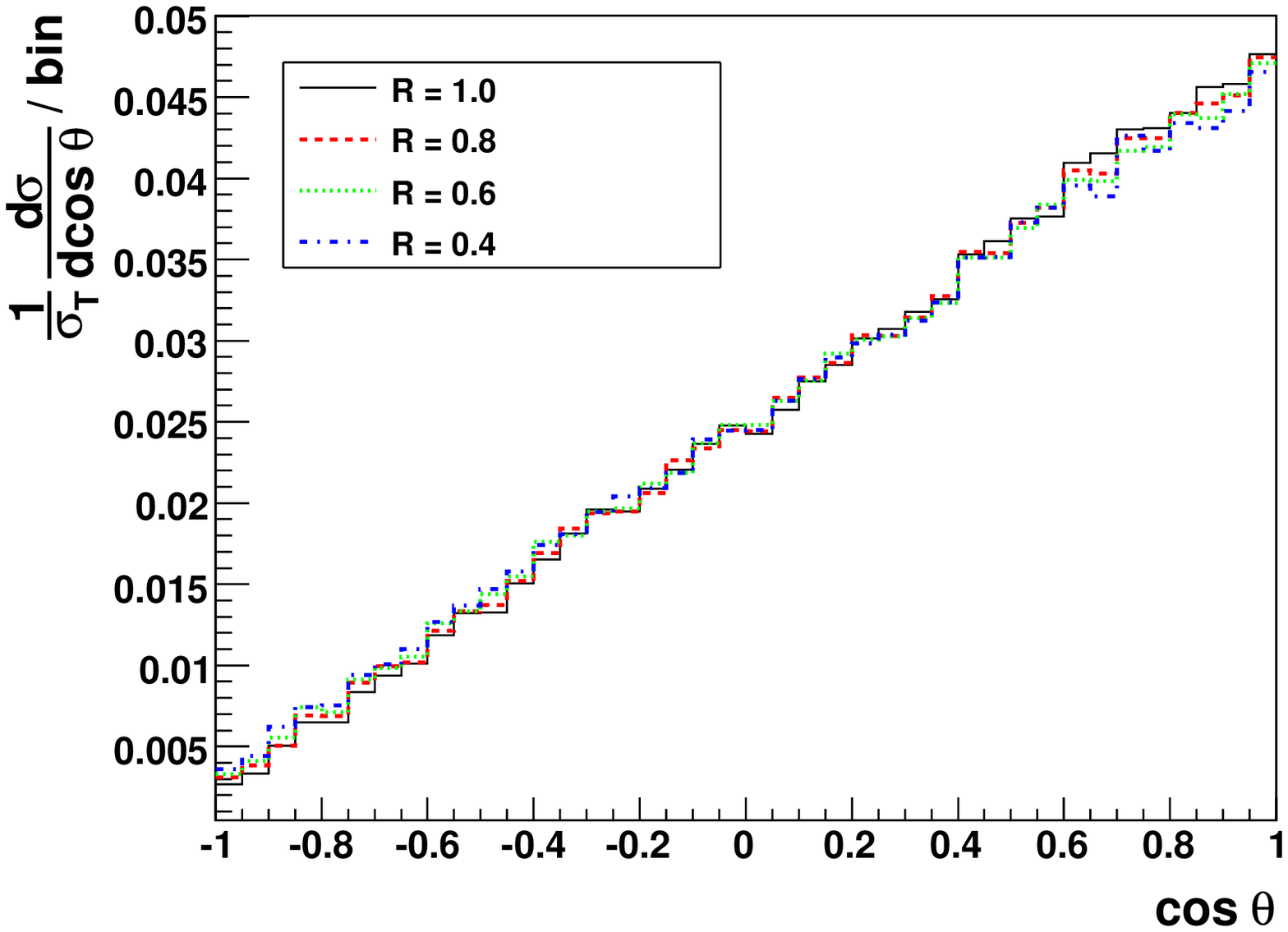}
  \includegraphics[width=0.49\textwidth]{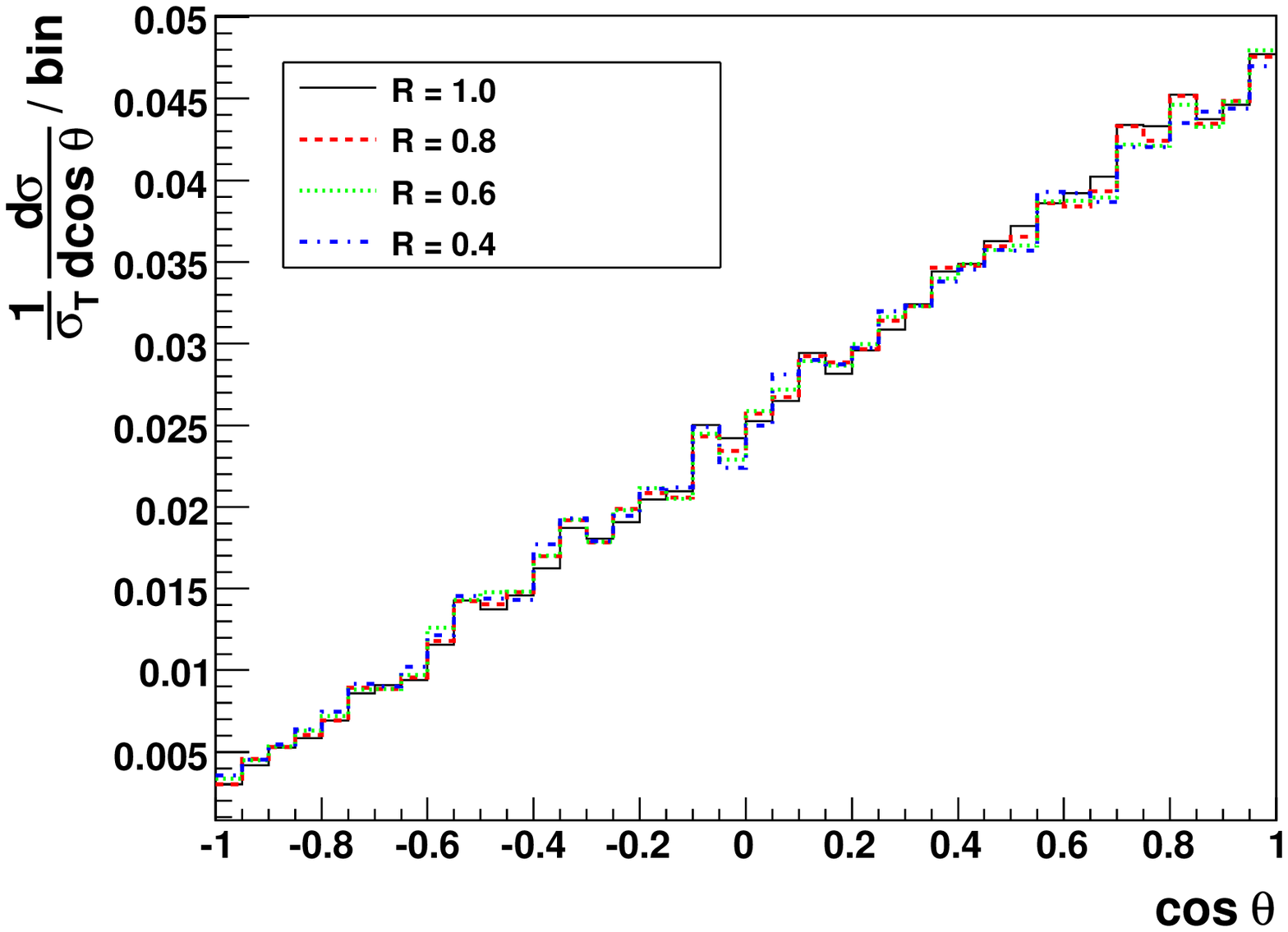}
  \caption{Angular correlation for four different values of $R$. The left plot is for the spectator basis while the right plot for the beamline basis with $|\eta_{j_1}|>$2.5.}
  \label{fig:Rs}
\end{figure}

\noindent The hardest (non-$b$) jet in the final state is likely to coincide with the spectator jet most of the times. In Fig.~\ref{fig:jet1-rap} the pseudorapidity $\eta$ of the hardest (non-$b$) jet is
shown for top and anti-top production, respectively. 
For both the top and anti-top production we see that
$d\sigma/d \eta$ peaks considerably around $|\eta|\sim 3$, corresponding to
an angle of about 5$^{\circ}$. In the case of single anti-top production we observe a somewhat larger distribution around $\eta=0$ than in the single top case. This means that there are more events where the spectator jet has more transverse momentum, which, in turn, suggests that the deflection away from the direction of the beamline is more probable. Therefore single anti-top quarks will be slightly less polarized in the spectator basis than single top quarks. Below we find that this is, in fact, the case. The plots in Fig.~\ref{fig:jet1-rap} also suggest that placing a cut on $|\eta_{j_1}|>\eta_{\textrm{cut}}=2.5$ will lead to an improvement in the degree of polarization for the beamline basis since the least forward of the jets will be left out.\\
In Fig.~\ref{fig:correl1} we show the angular correlations for the spectator basis, the beamline basis and the beamline basis with $|\eta_{j_1}|>2.5$ (the dotted line). The angle $\theta$ is defined as for Eq.(\ref{eq:decay1}). 
Although the differences appear to be small, we see that the correlations are strongest for the spectator basis and the beamline basis with a $\eta_{j_1}$-cut as compared to the beamline basis without cuts. In order to quantify the differences we use a linear fit to extract the polarization. The top decay distribution is given in Eq.(\ref{eq:decay1}) and we thus see that a linear fit of the form $a u + b$, where $u=\cos \theta$, to the distribution will yield the parameters $a$ and $b$. According to Eq.(\ref{eq:decay1}) we must have $a=s \mathcal{A}_{\uparrow \downarrow} =s (N_{\uparrow}-N_{\downarrow})/(N_{\uparrow}+N_{\downarrow})$, i.e. the spin asymmetry times the correlation factor. Using the spin asymmetry, and assuming $s=1$, it is possible to obtain the degrees of polarization\footnote{In fact it has been assumed that $s=1$ for all the numerical results presented in Tables~\ref{tab:table11}-\ref{tab:asym_bml_R}.}:
\begin{eqnarray}
  \label{eq:polar1}
  P_{t}=&\frac{1}{2}\left(s\mathcal{A}_{\uparrow \downarrow}+1 \right) \rvert_{s=1}&=\frac{1}{2}\left(\frac{N_{\uparrow}-N_{\downarrow}}{N_{\uparrow}+N_{\downarrow}} +1 \right) = \frac{N_{\uparrow}}{N_{\uparrow}+N_{\downarrow}}, \\
  P_{\bar{t}}=&\frac{1}{2}\left(-s\mathcal{A}_{\uparrow \downarrow}+1 \right) \rvert_{s=1}&=\frac{1}{2}\left(\frac{N_{\downarrow}-N_{\uparrow}}{N_{\uparrow}+N_{\downarrow}} +1 \right) = \frac{N_{\downarrow}}{N_{\uparrow}+N_{\downarrow}}. 
\end{eqnarray}
As a third measure we define the asymmetry (as done in~\cite{Stelzer:1998ni}):
\begin{equation}
  \label{eq:asym1}  A_{t/\bar{t}}=\frac{\sigma(\alpha \leq \cos \theta^{t/\bar{t}}_{l} < \beta ) - \sigma(\beta \leq \cos \theta^{t/\bar{t}}_{l} < \gamma )}{\sigma(\alpha \leq \cos \theta^{t/\bar{t}}_{l} < \beta ) + \sigma(\beta \leq \cos \theta^{t/\bar{t}}_{l} < \gamma )}
\end{equation}
where $\alpha, \beta, \gamma \in [-1:1]$ and $\alpha < \beta < \gamma$. For
the case without cuts we use $\alpha = -1$, $\beta = 0$ and $\gamma = 1$,
while for the cases where cuts have been applied we use $\alpha = -1$, $\beta
= -0.2$ and $\gamma = 0.6$. For the plots where kinematical cuts have been applied $A_{t/\bar{t}}$ will be the only (indirect) measure of the correlations. This is due to the fact that the cuts make the distributions less smooth, while for large $\theta$ there is a cut-off, which makes any attempt of fitting very empirical.\\ 
Using these definitions we summarize the numerical results in Tables~\ref{tab:table11} and~\ref{tab:table12}.
The spectator basis seems to be somewhat favored in the case of single top production while the beamline with an $\eta_{j_1}$-cut is better in the case of single anti-top production. This is thus in good accordance with what has been found in~\cite{Mahlon:1999gz} and the rapidity plots in Fig.~\ref{fig:jet1-rap}. \\
In order to make slightly more realistic predictions we next consider the following cuts (corresponding to the cuts in~\cite{Mahlon:1999gz}):\\
\begin{center}
\begin{tabular}[t]{l l l}
  $b$-cuts: & $|\eta_{b}| < 2.5$,  & $p_{T,b} > 50 \textrm{GeV}$, \\
 charged lepton cuts: &$|\eta_{l}| < 2.5$,  & $p_{T,l} > 20 \textrm{GeV}$, \\
 spectator jet cuts: &$2.5<|\eta_{j_1}| < 5$,  & $p_{T,j_1} > 50 \textrm{GeV}$, \\
 other cuts: & & $p_{T,\nu} > 20 \textrm{GeV}$. \\
\end{tabular}
\end{center}\mbox{}\\
As can be seen in Fig.~\ref{fig:correl2} the cuts reduce the cross section at very small angles, i.e. for $\cos \theta \rightarrow 1$.
The corresponding asymmetries are calculated using the formula in Eq.(\ref{eq:asym1}) and listed in Table~\ref{tab:tab2}.\\
\\
We now proceed with the second part. The only numerical results we present here are the top polarizations, i.e. we do not present asymmetries nor spin asymmetries. In each scenario we consider both the spectator basis and the beamline basis with $|\eta_{j_1}|>$2.5.\\
\begin{figure}[t!]
  \centering
  \includegraphics[width=0.49\textwidth]{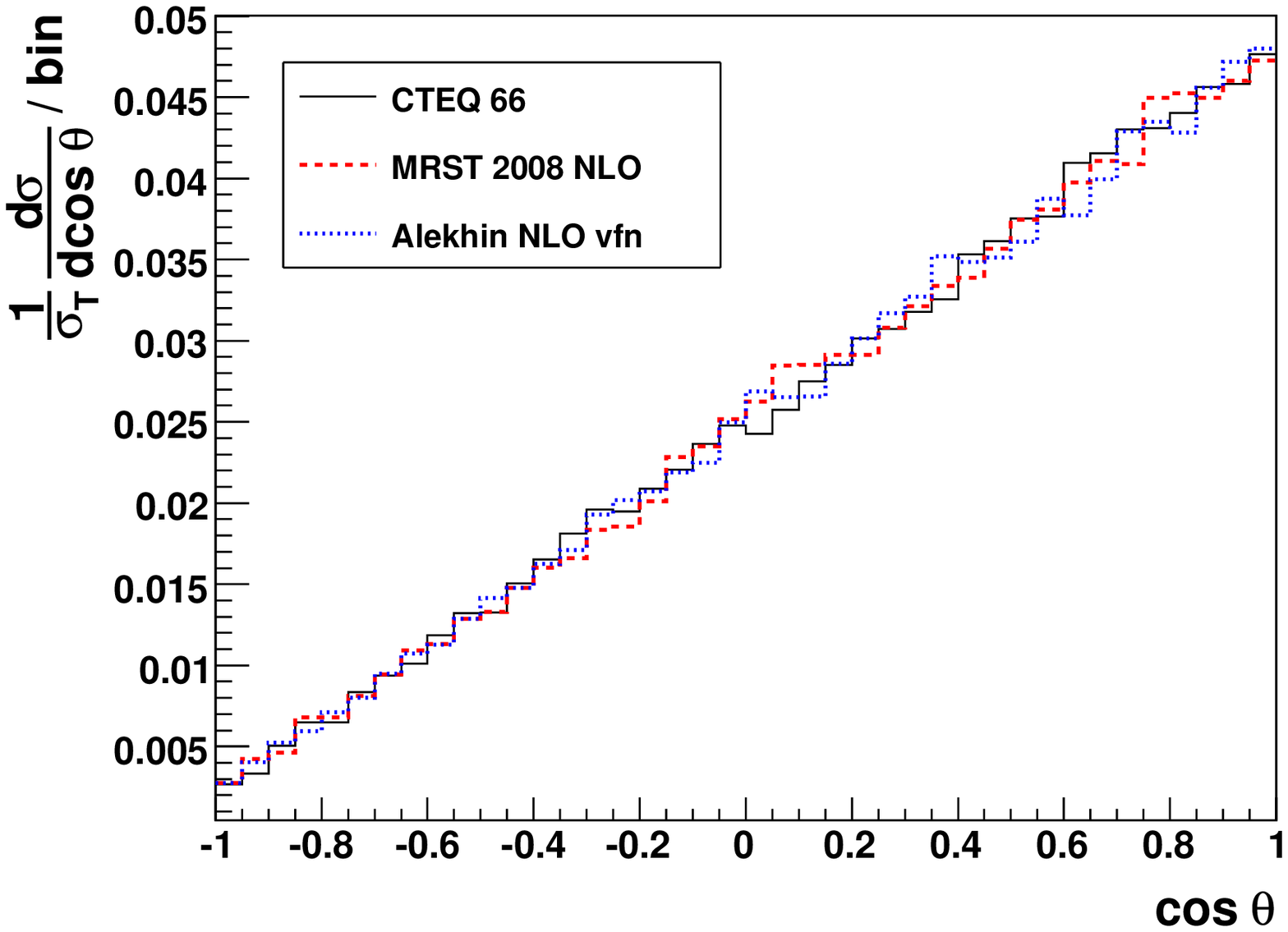}
  \includegraphics[width=0.49\textwidth]{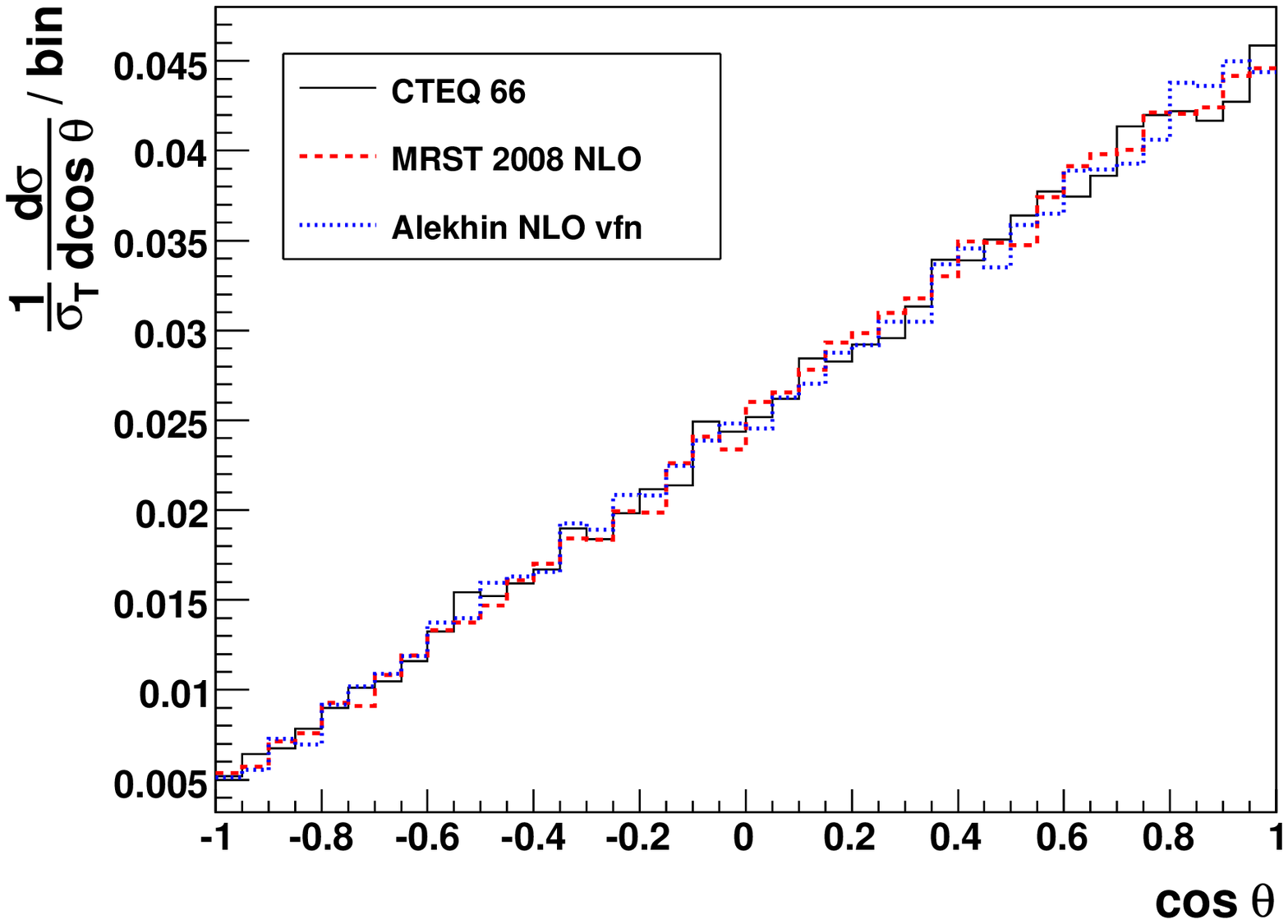}
  \caption{Angular correlation in single-$t$ production using three different PDF sets. The left plot is made using the spectator basis while the right plot using the beamline basis with $|\eta_{j_1}|>$2.5.}
  \label{fig:correl_pdf_spec}
\end{figure}
We begin with examining how much the choice of a particular PDF set affects the polarization. For this purpose we compare results from runs with the CTEQ66 (default), the MSTW2008nlo~\cite{Martin:2009iq} and the Alekhin vfn~\cite{Alekhin:2002fv} sets (all NLO). All other parameters are the same as in the first part. The results are shown in Fig.~\ref{fig:correl_pdf_spec}.
The figures reveal no particular difference which is confirmed by the corresponding numbers in Table~\ref{tab:asym_spec_pdf}.
In the beamline case with $|\eta_{j_1}|>$2.5 there are no qualitative differences between the three sets and for the spectator basis we see that the differences between the three sets lie well within one percent.\\
The dependence of the top polarization on the factorization scale, $\mu_F$, for each of the PDF sets is shown in Table~\ref{tab:mu_f}. We see that the polarization is almost stable with respect to $\mu_F$. For a given PDF set we see that the difference in polarization for the three different values of $\mu_F$ is contained well within 1\%, both for the spectator basis and the beamline basis with $|\eta_{j_1}|>$2.5.\\
\begin{figure}[t!]
  \centering
  \includegraphics[width=0.49\textwidth]{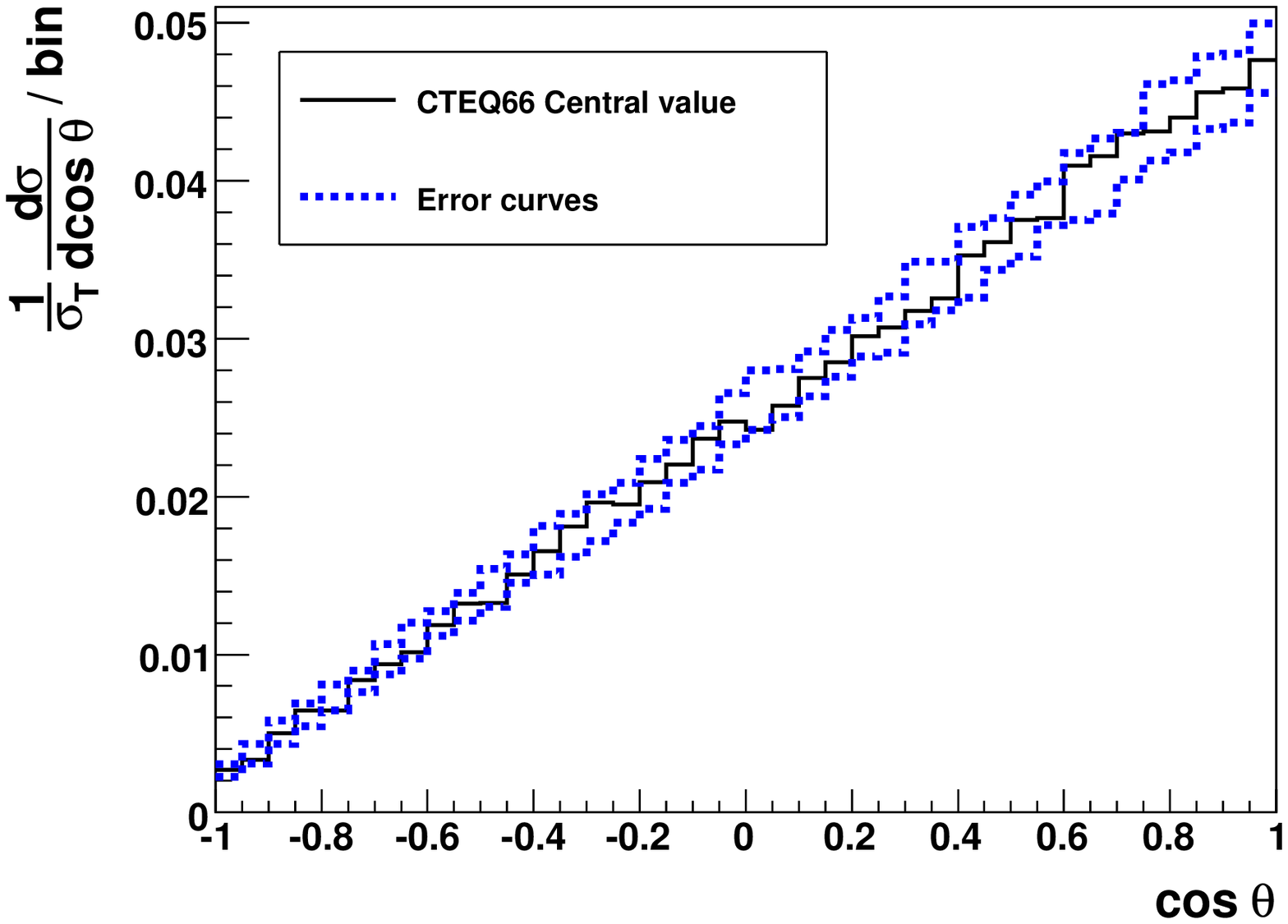}
  \includegraphics[width=0.49\textwidth]{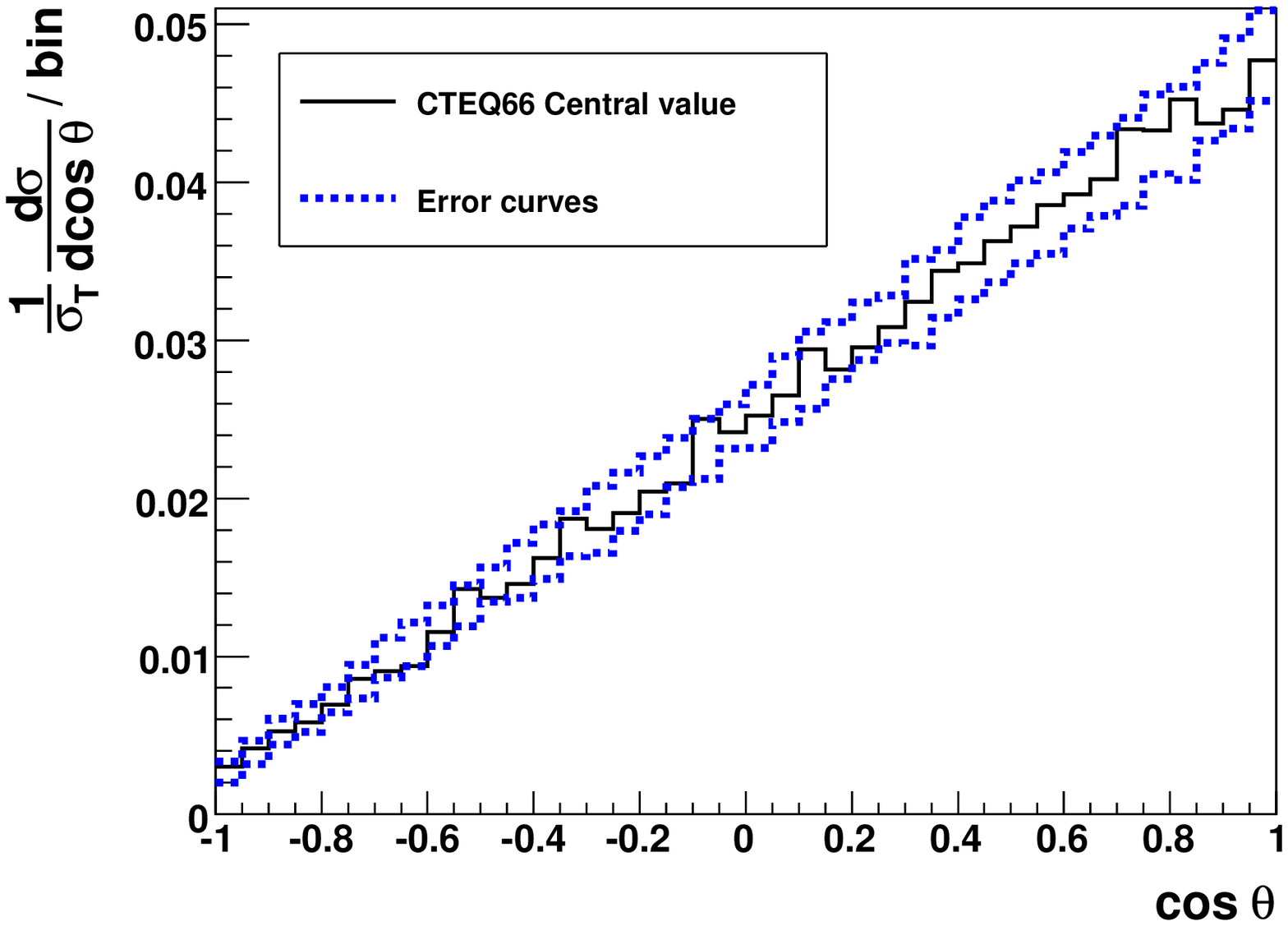}
  \caption{Figures showing the PDF error bands for the CTEQ66 PDF set. The left plot is for the spectator basis while the right plot is for the beamline basis with $|\eta_{j_1}|>$2.5.}
  \label{fig:PDF_err}
\end{figure}
In Fig.~\ref{fig:PDF_err} the PDF error bands for the default (CTEQ66) PDF set are shown. The slopes of both the upper- and lower boundary are different from the slope of the curve corresponding to the central value, leading to a cone-shaped error band. The different slopes are used to determine the uncertainty of the polarization due to the PDF errors. In Table~\ref{tab:pdf_err} the default (CTEQ66) polarizations are shown with the corresponding PDF errors which are seen to be contained within 3\%. 
\\
Until now all the results presented in this letter are based on the assumption that the center-of-mass energy of the colliding hadrons is 14 TeV. Experiments at the LHC will, however, begin at lower energies~\cite{LHCcommission}. It is thus interesting to find out how the angular correlations depend on the CM energy. In Fig.~\ref{fig:ECM} we show the correlations for $\sqrt{S}=$14 TeV (default), $\sqrt{S}=$10 TeV and $\sqrt{S}=$5 TeV respectively.
The correlation curves reveal no visible variation with respect to $\sqrt{S}$ which is supported by the numbers in Table~\ref{tab:asym_spec_ECM}. 
We conclude that the correlations are stable with respect to the center of mass energy. This is to be expected since $\sqrt{S}=5$ TeV is already well above production threshold.\\
\\
In establishing the spectator jets we used the $k_T$-clustering jet algorithm. It is possible to define a so-called jet radius, $R$, as done in~\cite{Seymour:1994by} i.e.:
\begin{equation}
  \label{eq:jetradius}
  d_i=p_{T,i}^2 R^2
\end{equation}
which is similar to the radius of the jet cone in the cone algorithm. Since the choice of spin quantization axis of the top depends crucially on the spectator jet kinematics, regardless whether the spectator or the beamline basis is chosen, we investigate how the correlation behaves with respect to the $R$--parameter. This is shown in Fig.~\ref{fig:Rs}.
The numbers (see Table~\ref{tab:asym_bml_R}) show the same trend for both the spectator basis and the beamline basis: the polarization decreases when the radius becomes smaller. 
However the decrease in polarization is small despite a rather large variation of $R\in[0.4;1]$ and hence the correlations seem robust with respect to $R$.

\section{Conclusions}
\label{sec:concl}

Using the MC@NLO framework we have investigated the prospects of observing angular correlations\footnote{We have not estimated statistical errors in this paper.} in the $t$-channel of single top and anti-top production at the LHC. This is the first study, to our knowledge, done using an event generator accurate to NLO and matched appropriately to MC parton showers. \\
We have showed, in accordance with~\cite{Mahlon:1999gz}, that both the spectator basis and the beamline basis define ensembles of highly polarized top quarks in the $t$-channel. Comparing single top and anti-top production in the $t$-channel we found that for single top production and decay the correlations are strongest when using the spectator basis and the beamline basis with $|\eta_{j_1}|>$2.5. For single anti-top production and decay the beamline basis with the $\eta_{j_1}$-cut is slightly the better choice leading to the highest degree of polarization. The overall qualitative differences are small, however, and contained within a couple of percent.\\
In this letter we have illustrated that the angular correlations and amount of top polarization are largely unaffected by changing the PDF set and by a lowering of the center-of-mass energy, provided the energy is well above production threshold. The dependence on the factorization scale was shown to be weak and the PDF errors were shown to lead to uncertainties of less than 3\% for the polarization. Also, we investigated the impact it has on the angular correlations when the $R$--parameter is changed in the $k_T$-jet clustering algorithm. We found that the correlations became weaker when $R$ was diminished. It should, however, be pointed out that despite rather large variations of $R$ the correlations remained strong. We thus conclude that the correlations are robust with respect to the $R$--parameter.   

\subsection*{Acknowledgements}
\label{acknow}

The author would like to thank W. Bernreuther, S. Dittmaier and E. Laenen for fruitful discussions and for their suggestions.

\newpage

\section*{Tables}
\label{tables}
\begin{table}[h]
\caption{Asymmetries, spin-asymmetries, and degrees of polarization for $t$-channel single top production corresponding to three choices of spin quantization bases. (see left plot in Fig.~\ref{fig:correl1}.)}
\centering
\begin{tabular}{l|c|c|c|}
  \textbf{Basis} & $A_{t}$ & $\left(\frac{N_{\uparrow}-N_{\downarrow}}{N_{\uparrow}+N_{\downarrow}} \right)_t$ & $P_t$ \\ 
  \hline
  Spectator & -0.459  & 0.923 $\pm$ 0.006  & 96.16\% $\pm$ 0.30\% $\uparrow$  \\
  $\eta$-bml & -0.414  & 0.823 $\pm$ 0.006  & 91.15\% $\pm$ 0.32\% $\uparrow$ \\
  $\eta$-bml w. $\eta$-cut & -0.462 & 0.914 $\pm$ 0.008  & 95.68\% $\pm$ 0.38\% $\uparrow$  \\
  \hline
\end{tabular}
\label{tab:table11}
\end{table}
\begin{table}[h]
\caption{Asymmetries, spin-asymmetries, and degrees of polarization for $t$-channel single anti-top production corresponding to three choices of spin quantization bases. (see right plot in Fig.~\ref{fig:correl1}).}
\centering
\begin{tabular}{l|c|c|c|}
  \textbf{Basis} & $A_{\bar{t}}$ & $\left( \frac{N_{\uparrow}-N_{\downarrow}}{N_{\uparrow}+N_{\downarrow}} \right)_{\bar{t}}$ &
 $P_{\bar{t}}$ \\
  \hline

  Spectator & -0.438   & -0.882 $\pm$0.006  & 94.10\% $\pm$ 0.32\% $\downarrow$ \\
  $\eta$-bml & -0.420  & -0.837 $\pm$0.007 & 91.87\% $\pm$ 0.33\% $\downarrow$ \\
  $\eta$-bml w. $\eta$-cut & -0.470  & -0.936 $\pm$ 0.008  & 96.82\% $\pm$ 0.41\% $\downarrow$ \\
  \hline
\end{tabular}
\label{tab:table12}
\end{table}
\begin{table}[h]
\caption{Asymmetries of the plots in Fig.~\ref{fig:correl2}.}
\centering
\begin{tabular}{l|c||c|}
  \textbf{Basis} & $A_{t}$ & $A_{\bar{t}}$ \\
  \hline
  Spectator  & -0.608 & -0.572 \\
  $\eta$-bml & -0.583 & -0.570 \\
  \hline
\end{tabular}
\label{tab:tab2}
\end{table}
\begin{table}[h!]
\caption{Degrees of polarization corresponding to three different choices of PDF sets.}
\centering
\begin{tabular}{l|c|c|}
  \textbf{PDF set} & \textbf{Spectator basis} & \textbf{Beamline basis w. $|\eta_{j_1}|>$2.5} \\
  \hline
  CTEQ66 & 96.16\% $\pm$ 0.30\% $\uparrow$ & 91.15\% $\pm$ 0.32\% $\uparrow$ \\
  MSTW2008nlo & 95.28\% $\pm$ 0.38\% $\uparrow$ & 91.01\% $\pm$ 0.40\% $\uparrow$ \\
  ALEKHIN vfn & 95.81\% $\pm$ 0.30\% $\uparrow$ & 91.36\% $\pm$ 0.32\% $\uparrow$\\
  \hline
\end{tabular}
\label{tab:asym_spec_pdf}
\end{table}
\newpage
\begin{table}[h]
\caption{Degrees of polarization corresponding to three different PDF sets and three different values of the factorization scale.}
\centering
 \textbf{Spectator basis}
\begin{tabular}{l|c|c|c|}
  $\mu_F/m_t$ & CTEQ66 & MSTW2008nlo & AlekhinNLOvfn \\
  \hline
  0.5 & 95.34\% $\pm$ 0.30\% $\uparrow$ & 95.75\% $\pm$ 0.32\% $\uparrow$ & 95.84\% $\pm$ 0.30\% $\uparrow$ \\
  1 & 96.16\% $\pm$ 0.30\% $\uparrow$ & 95.91\% $\pm$ 0.31\% $\uparrow$ & 95.81\% $\pm$ 0.30\% $\uparrow$ \\
  2 & 96.11\% $\pm$ 0.30\% $\uparrow$ & 96.83\% $\pm$ 0.34\% $\uparrow$ & 96.21\% $\pm$ 0.30\% $\uparrow$ \\
  \hline
\end{tabular}\\
\mbox{}\\
\mbox{}\\
\textbf{Beamline basis w. $|\eta_{j_1}|>$2.5}
\begin{tabular}{l|c|c|c|}
  $\mu_F/m_t$ & CTEQ66 & MSTW2008nlo & AlekhinNLOvfn \\
  \hline
  0.5 & 95.33\% $\pm$ 0.39\% $\uparrow$ & 95.39\% $\pm$ 0.40\% $\uparrow$ & 95.42\% $\pm$ 0.38\% $\uparrow$ \\
  1 & 95.68\% $\pm$ 0.38\% $\uparrow$ & 95.72\% $\pm$ 0.39\% $\uparrow$ & 95.59\% $\pm$ 0.38\% $\uparrow$\\
  2 & 95.75\% $\pm$ 0.38\% $\uparrow$ & 96.50\% $\pm$ 0.42\% $\uparrow$ & 96.13\% $\pm$ 0.38\% $\uparrow$ \\
  \hline
\end{tabular}
\label{tab:mu_f}
\end{table}
\begin{table}[h!]
\caption{Degrees of polarization corresponding to the central value (default) with PDF errors.}
\centering
\begin{tabular}{l|c|c|}
   & \textbf{Spectator basis} & \textbf{Beamline basis w. $|\eta_{j_1}|>$2.5} \\
  \hline
  CTEQ66 & $\textrm{96.16}^{+\textrm{1.70}}_{-\textrm{2.03}}$ \% $\pm$ 0.30\% $\uparrow$ & $\textrm{95.68}^{+\textrm{2.46}}_{-\textrm{2.19}}$\% $\pm$ 0.38\% $\uparrow$ \\
  \hline
\end{tabular}
\label{tab:pdf_err}
\end{table}
\begin{table}[h]
\caption{Degrees of polarization corresponding to three different center of mass energies, $\sqrt{S}$.}
\centering
\begin{tabular}{l|c|c|}
  \textbf{$\sqrt{S}$ (TeV)} & \textbf{Spectator basis} & \textbf{Beamline basis w. $|\eta_{j_1}|>$2.5} \\ 
  \hline
  14 & 96.16\% $\pm$ 0.30\% $\uparrow$ & 95.68\% $\pm$ 0.38\% $\uparrow$ \\
  10 & 96.02\% $\pm$ 0.28\% $\uparrow$ & 95.99\% $\pm$ 0.36\% $\uparrow$ \\
  5 & 95.98\% $\pm$ 0.24\% $\uparrow$ & 96.48\% $\pm$ 0.36\% $\uparrow$  \\
  \hline
\end{tabular}
\label{tab:asym_spec_ECM}
\end{table}
\begin{table}[h]
\caption{Degrees of polarization corresponding to four different values of the jet $R$--parameter.}
\centering
\begin{tabular}{l|c|c|c|c|}
  \textbf{$R$} & \textbf{Spectator basis} & \textbf{Beamline basis w. $|\eta_{j_1}|>$2.5} \\
  \hline
  1.0 & 96.16\% $\pm$ 0.30\% $\uparrow$ & 95.68\% $\pm$ 0.38\% $\uparrow$ \\
  0.8 & 95.27\% $\pm$ 0.28\% $\uparrow$ & 95.31\% $\pm$ 0.36\% $\uparrow$ \\
  0.6 & 94.51\% $\pm$ 0.26\% $\uparrow$ & 94.88\% $\pm$ 0.34\% $\uparrow$ \\
  0.4 & 93.73\% $\pm$ 0.25\% $\uparrow$ & 94.55\% $\pm$ 0.33\% $\uparrow$ \\
  \hline
\end{tabular}
\label{tab:asym_bml_R}
\end{table}

\end{document}